\documentclass[aps,prb,reprint,groupedaddress]{revtex4-1}
\usepackage{amsmath,amsthm,amssymb,bm}
\usepackage[dvipdfmx]{graphicx}

\begin{document}


\title{Site-specific $^{13}$C NMR study on locally distorted triangular lattice of organic conductor $\kappa$-(BEDT-TTF)$_2$Cu$_2$(CN)$_3$}


\author{Y. Saito}
\author{T. Minamidate}
\author{A. Kawamoto}
\email[]{atkawa@phys.sci.hokudai.ac.jp}
\author{N. Matsunaga}
\author{K. Nomura}
\affiliation{Department of Physics, Hokkaido University, Sapporo 060-0810, Japan}


\date{\today}

\begin{abstract}
To verify the effect of geometrical frustration, we artificially distort the triangular lattice of quasi-two-dimensional organic conductor $\kappa$-(BEDT-TTF)$_2$Cu$_2$(CN)$_3$ [BEDT-TTF: bis(ethylenedithio)terathiofulvalene] by analogous-molecular substitution and apply $^{13}$C NMR of bulk and substituted sites, electric conductivity, and static magnetic susceptibility measurements. The results indicate that the magnetic characteristics of the substituted sample are quantitatively similar to those of the pure sample. Moreover the magnetic characteristics at the substituted sites are also the same as in the bulk. These results suggest that the observed magnetic properties may not be due to the geometrical frustration but the importance of disorder.
\end{abstract}


\maketitle

\section{Introduction}
$\kappa$-(BEDT-TTF)$_2X$ salts are well-known organic conductors based on bis(ethylenedithio)tetrathiafulvalene (BEDT-TTF) molecules which form dimers in the conduction layer. The compound $X^-$ is a monovalent anion, so the formal charge of BEDT-TTF is +0.5. Although this is regarded as a quarter-filled system, the electronic state of $\kappa$-(BEDT-TTF)$_2X$ is believed to be half filled due to the dimerization of the BEDT-TTF molecules \cite{Kanoda1997}.

At ambient pressure, the ground state of $\kappa$-(BEDT-TTF)$_2$Cu$_2$(CN)$_3$ [hereafter abbreviated as the (CN)$_3$ salt] is expected to be antiferromagnetic (AF) because its ratio of $U/W$ is greater than that for $\kappa$-(BEDT-TTF)$_2$Cu[N(CN)$_2$]Cl [hereafter abbreviated as the Cl salt] \cite{Komatsu1996}, where $U$ is the effective onsite Coulomb repulsion and $W$ is the bandwidth, respectively. Contrary to the expectation, the (CN)$_3$ salt exhibits no magnetic ordering \cite{Shimizu2003}. One possible reason no-AF ordering occurs in (CN)$_3$ salt is the contribution of geometrical frustration of spins \cite{Shimizu2003}. From density functional theory, we know that at room temperature, the ratio of interdimer transfer integrals $t'/t$ = 0.83 is very close to unity \cite{Kandpal2009,Nakamura2009,Jeschke2012}, where $t$ is the nearest neighbor transfer and $t'$ is the second-nearest neighbor transfer, respectively. Moreover, as the exchange interaction, $J(t)= -2t^2/U$, the ratio of $J(t')/J(t)$, is close to unity, suggesting a nearly isotropic triangular lattice with $S = 1/2$. In this case, AF ordering could be suppressed by the geometrical frustration of the spins \cite{Anderson1973}. However, optical conductivity measurements of (CN)$_3$ salt do not reveal a clear-cut energy gap at all temperatures \cite{Kezsmarki2006,Elsasser2012}, which is expected from a Mott insulator due to the concomitant large $U/W$. Moreover, charge instability below 60 K was reported in (CN)$_3$ salt by $^{13}$C nuclear magnetic resonance (NMR) \cite{Kawamoto2004, Kawamoto2006} and Raman spectroscopy \cite{Yakushi2015}.

Disorder or randomness is a source of localization in strongly correlated electron systems. In this case, conduction electrons localize by impurity scattering to form an \textit{Anderson insulator}. Contrary to a Mott insulator, no gap appears at the Fermi surface in Anderson insulators. Some theories predict that a soft Hubbard gap whose density of state with zero density of states at $E_{\rm F}$ emerges by introducing disorder to the strongly correlated electron system \cite{Byczuk2005,Aguiar2009,Shinaoka2009}. Due to competition between electron correlation and randomness, a Mott--Anderson transition is expected. Recently, the disorder in the anion groups has been suggested to act on the BEDT-TTF layers via hydrogen bonds, whereby the charge distribution is altered and domain boundaries appear in (CN)$_3$ salt \cite{Pinteric2014}. Therefore, the relationship between the physical properties and disorder in (CN)$_3$ salt is of significant research interest.

Experimental observations in (CN)$_3$ salt have been explained through triangular-lattice or disordered-state models. Therefore, experimental verification is required to determine whether the observed behavior is due to the the geometrical frustration of the spins in the triangular lattice or to a disordered state. One approach is to artificially distort triangular lattices. To investigate the disorder effects of the Mott-insulating state, Cl salt was irradiated by x-ray irradiation, which introduces disorder and changes the Mott-insulating state of the Cl salt into a soft-Hubbard-gapped insulating state \cite{Sasaki2012} and, simultaneously, AF ordering disappears \cite{Furukawa2015}.

\begin{figure}[htbp]
  \includegraphics[width=5cm]{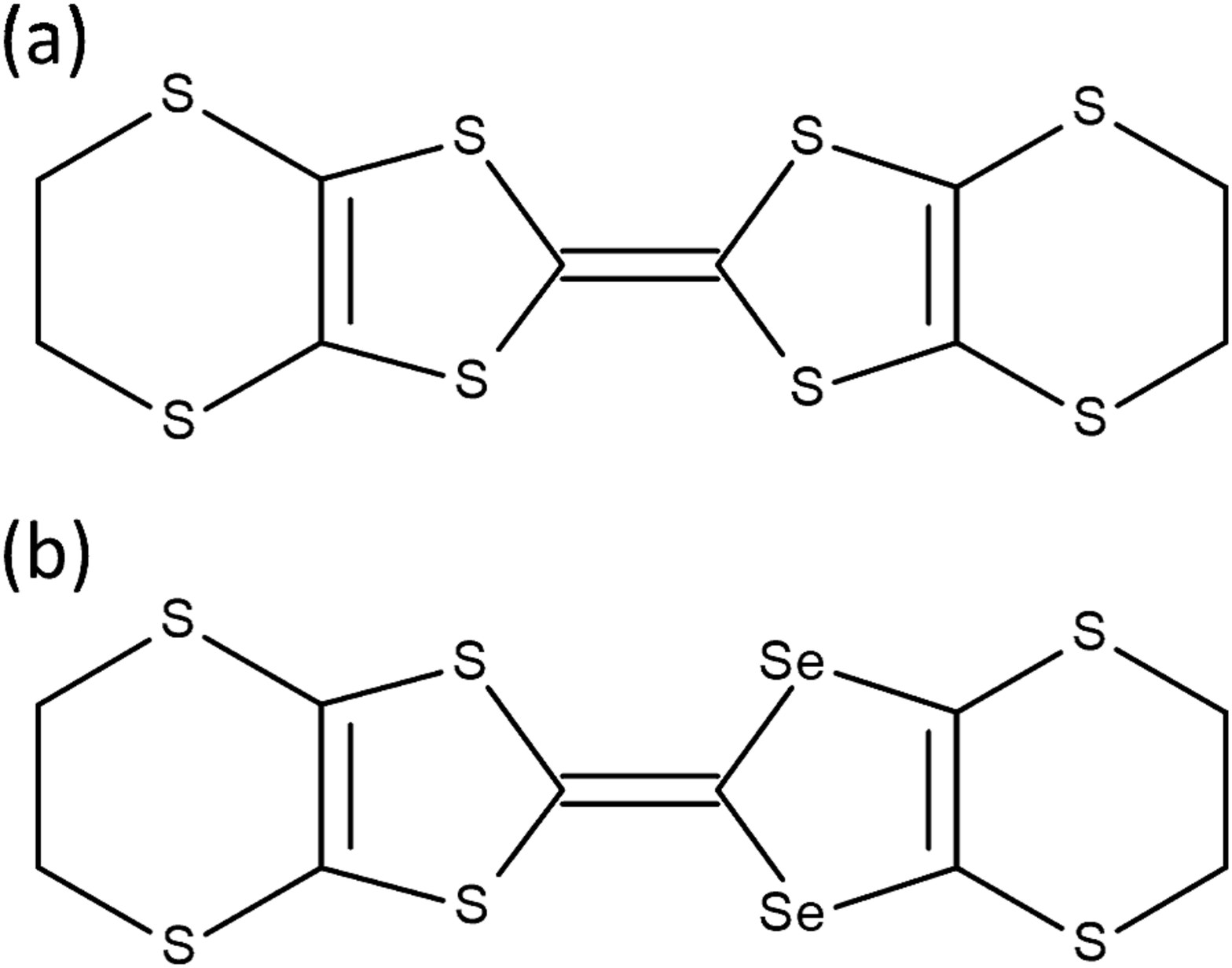}
  \caption{Molecular structures of (a) BEDT-TTF and (b) us-BEDT-STF.}\label{molecule}
\end{figure}

Note that x-ray irradiation mainly affects on the crystal surface owing to the absorption by Cu atoms and produces free radicals on the surface. Therefore, this technique is not suited for investigating the magnetic properties. Moreover, it is unclear whether x-ray irradiation and anion disorder via hydrogen bonding in pure samples introduces disorder in the conduction plane. An alternative approach is analogous-molecular substitution to directly introduce disorder into the conduction layer, which is the method we use herein. Specifically, we focus on a unsymmetric (us-) bis-(ethylenedithio)diselenadithiafulvalene (BEDT-STF), where in one side of the S in the central ring of the TTF skeleton is replaced with Se atoms (see Fig. \ref{molecule} (b)). The advantages of substituting BEDT-STF for BEDT-TTF are that (i) structural distortion is minimized because BEDT-TTF and BEDT-STF have almost the same molecular structure, (ii) the large spread of Se 4$d$ orbital creates sufficient disorder to affect the intermolecular transfer integrals, and (iii) the substitution fraction and uniformity can be determined by elemental analysis to detect the Se atoms, which exist only at substituted sites.
However, the primary important advantage is that the transfer integrals around the BEDT-STF molecule are modified by the Se orbitals, which locally distorts the triangular lattice.

NMR is a microscopic probe that measures the static and dynamic magnetic properties. The local spin susceptibility can be determined from the Knight shift $K$, and magnetic fluctuations from the spin-lattice relaxation rate $T_1^{-1}$. Many previous NMR studies of BEDT-TTF salts have contributed important information on these salts \cite{Kagawa2009,Mayaffre2014}. Herein, to elucidate the magnetic properties of (CN)$_3$ salt, we apply $^{13}$C NMR to bulk BEDT-TTF sites and compare the NMR results of pure samples with those of BEDT-STF-substituted samples.

Moreover, it is expected that local impurities induce staggered moments at the center of disorder in the vicinity of antiferromagnetic compounds. However, few NMR studies of impurity sites exist. To address this void, we substitute $^{13}$C enriched BEDT-STF for non-enriched BEDT-TTF molecules so that $^{13}$C-NMR targets the impurity sites. 

\section{Experimental}

Single crystals of $\kappa$-[(BEDT-TTF)$_{1-x}$(BEDT-STF)$_x$]$_2$Cu$_2$(CN)$_3$ of various stoichiometry ($x=0$, 0.05, and 0.06) were prepared by the electrochemical oxidation \cite{Geiser1991}. Using energy-dispersive x-ray spectroscopy with $\lambda$-(BEDT-STF)$_2$GaCl$_4$ as a reference, the impurity fraction $x$ was determined by comparing the intensity of S atoms to that of Se atoms. Samples with $x$=0 and 0.05 consisted of $^{13}$C-enriched BEDT-TTF and non-enriched BEDT-STF, and a sample with $x=0.06$ consisted of $^{13}$C-enriched BEDT-STF and non-enriched BEDT-TTF. To avoid the Pake-doublet problem \cite{Pake1948}, one side of the central C=C bond in BEDT-TTF, and that of the central C=C bond far from Se atoms in BEDT-STF are enriched with the $^{13}$C isotope. $^{13}$C-substituted molecule were prepared through cross coupling \cite{Hirose2012}. DC conductivity was measured along the $b$ axis from room temperature down to 40 K for the pure and $x=0.05$ samples by the standard four-point probe technique. DC magnetization was measured for polycrystalline samples as a function of temperature from 300 to 2 K in a 2 T magnetic field using a magnetometer that is based on a superconducting quantum interference device. NMR experiments were performed for each single crystal in a 7 T magnetic field applied perpendicular to conduction plane. The NMR shifts are given in ppm relative to tetramethylsilane. The spin-lattice relaxation rate, $T^{-1}$, was measured by the conventional saturation-recovery method. The linewidths were determined by fitting peaks to a Lorentz function, and the spin-spin relaxation rate, $T_2^{-1}$, is defined as the rate corresponding to Lorentz decay.

\section{Results and discussion}
\subsection{Modification of transfer integrals}
\begin{figure}[htbp]
  \includegraphics[scale=0.25]{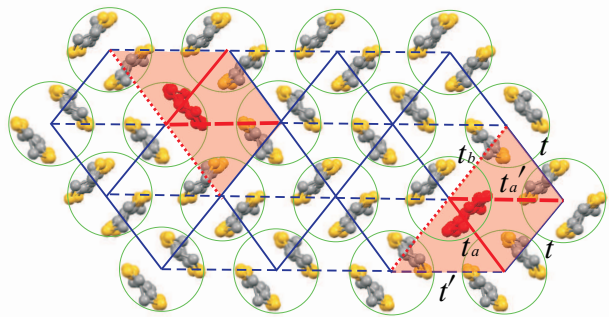}
  \caption{Schematic view of BEDT-TTF dimers in $bc$ plane. Red molecules represent BEDT-STF molecules. Original nearly isotropic and modified triangular lattices are shown by blue and red lines, respectively, where $t$ and $t'$ are the original transfer integrals and $t_a$, $t_a'$, and $t_b$ are the transfer integrals around the BEDT-STF molecule.}\label{suppressfrustration}
\end{figure}

We consider the modification of transfer integrals due to the extended 4$d$ orbital of the Se atoms. The 5 \%, BEDT-STF substitution shown in Fig. \ref{molecule}(b) modifies the transfer integrals of the affected sites as shown in Fig. \ref{suppressfrustration}. In fact, the transfer integral in the side-by-side direction of the BEDT molecules of $\alpha$-(BEDT-STF)$_2$I$_3$ is 46.5\% greater than that of $\alpha$-(BEDT-TTF)$_2$I$_3$ \cite{Inokuchi1995}. Therefore, the interdimer transfer integrals around the substituted dimer may be estimated to be approximately 20\% greater than that of the original dimer. As shown in Fig. \ref{suppressfrustration}, 20\% of the triangular lattices are distorted; thus, the resulting area of the original triangular lattice is confined to a radius of 2--3 dimers. To verify the effect of disorder, we measured the electrical conductivity and compared the results with those of previous reports.

\subsection{Conductivity}

\begin{figure}[htbp]
  \includegraphics[width=7.9cm]{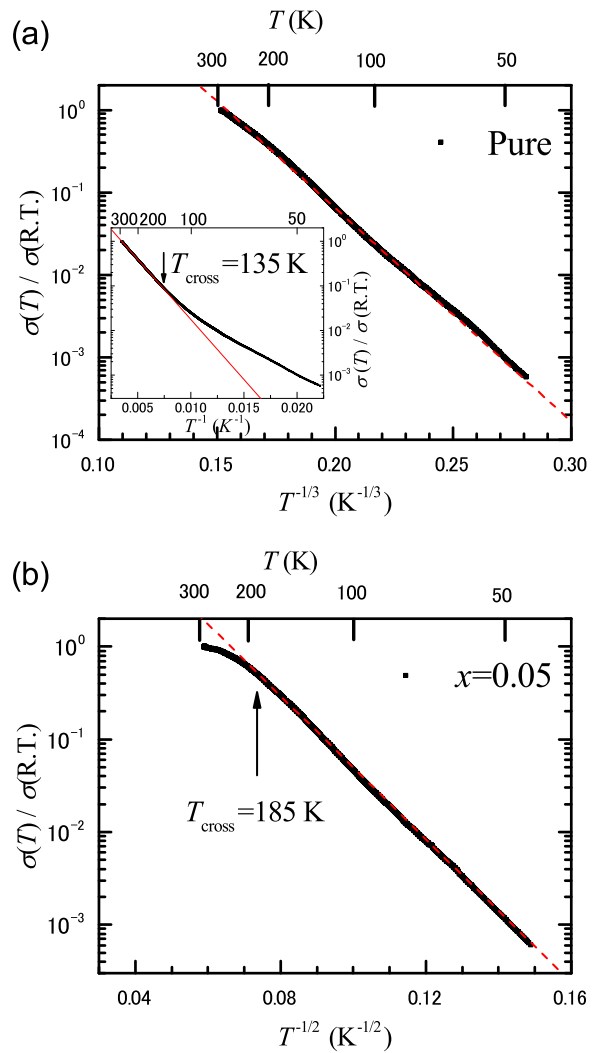}
  \caption{Normalized conductivity of (a) pure and (b) $x=0.05$ samples as a function of temperature. Inset shows Arrhenius plot. The red dashed (solid) line shows the fit given by the VRH (NNH) model.}\label{conductivity}
\end{figure}

Figure \ref{conductivity} shows conductivity normalized to the data at room temperature for $E \parallel b$ as a function of temperature for both samples. In the pure sample, the result is consistent with that of previous report \cite{Pinteric2014}, whereas the decrease of the conductivity near room temperature is suppressed in the $x = 0.05$ sample. The conductivities of both samples are fit by the nearest-neighbor-hopping (NNH) and variable-range-hopping (VRH) equations.
\begin{align}
 \sigma (T) & \propto \exp(-\Delta/T), \\
 \sigma (T) & \propto \exp[(-T_0/T)^{1/(d+1)}],
\end{align}
where $\Delta$ is the activation energy for the NNH model, $T_0$ is the activation energy for the VRH model, $d$ is the dimension of the VRH model, and $T_{\rm cross}$ is the crossover temperature from the NNH model in the high-temperature region to the VRH model in the low temperature region. For the pure sample, the conductivity can be fit by the NNH model above $T_{\rm cross}$, wheareas the VRH model with $d=2$ is suitable below $T_{\rm cross}$, as shown in Fig. \ref{conductivity} (a) where the dashed (solid) line represents the fit with the VRH (NNH) model (the fitting parameters are given in Table \ref{transport}).
For $x=0.05$ sample, the VRH model with $d=1$ is suitable compared to $d=2$ below $T_{\rm cross}$. Conductivity above $T_{\rm cross}$ significantly differs between the pure and $x=0.05$ samples. For $x=0.05$ sample, temperature dependence of conductivity is not much changed above $T_{\rm cross}$. Although the temperature dependence of the conductivity can be fit by the NNH model, the obtained value of $\Delta$ is too small to explain the crossover from the NNH to VRH model. The results indicate that the substitution enhances conduction and increases $T_{\rm cross}$ above that of the pure sample. As BEDT-STF substitution does not introduce carriers, we must rule out any effect due to carrier doping. To explain the results for $d=1$ and the $x=0.05$ sample, we suggest that competition between electron correlations and randomness possibly play a role \cite{Shinaoka2009}. For both samples, the VRH fit suggests that no intrinsic energy gap opens, and the remarkable effect of substitution was confirmed by studying the transport properties of the $x=0.05$ sample.
\begin{table}
 \begin{center}
  \caption{Transport parameters.}
   \begin{tabular}{|c|c|c|} \hline
         & Pure sample & $x=0.05$ sample \\ \hline
$\Delta$ (K) &  620    &  N/A    \\ \hline
$T_0$ (eV)   & 18.3    &0.68      \\ \hline
$T_c$ (K)   &   135   &   185   \\ \hline
$d$   &   2  &   1   \\ \hline
    \end{tabular}
    \label{transport}
  \end{center}
\end{table}
\subsection{Static magnetic susceptibility}

\begin{figure}[htbp]
  \includegraphics[width=7.9cm]{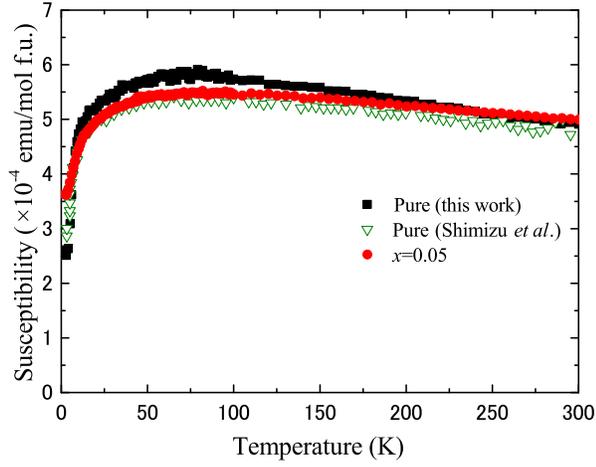}
  \caption{Magnetic susceptibility as a function of temperature for pure samples and sample with impurity fraction $x=0.05$.}\label{susceptibility}
\end{figure}

The static magnetic susceptibility of the pure and $x=0.05$ samples are shown in Fig. \ref{susceptibility}, along with the result of previous work \cite{Shimizu2003}. In this figure, the core diamagnetic contribution of $-4.37 \times 10^{-4}$ emu/mol f.u. \cite{Shimizu2003} is already subtracted. For $x=0.05$ sample, the temperature dependence is quantitatively fairly similar to that of the pure sample; both exhibit a hump around 60 K and rapidly decrease below 20 K. The value of $\chi = 5\times 10^{-4}$ emu/mol f.u. at 300 K is greater than that found for other $\kappa$ salts, i.e., $\chi = 4.5\times 10^{-4}$ emu/mol f.u. \cite{Kawamoto1995} The static susceptibility remains essentially unchanged on distortion of the triangular lattice.
\subsection{NMR spectra, line shift and linewidth}

\begin{figure}[htbp]
  \includegraphics[width=7.5cm]{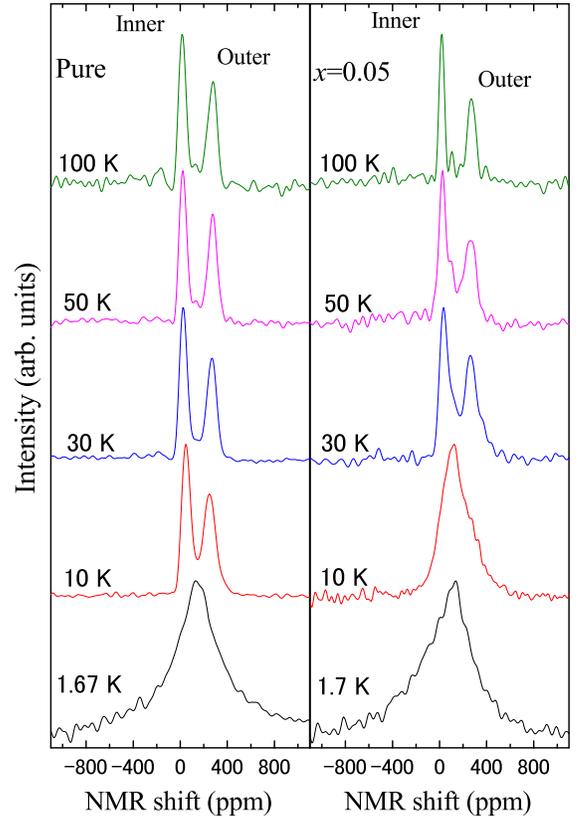}
  \caption{NMR spectra at several temperatures. Left (right) panel shows the results for the pure ($x=0.05$) sample.}\label{spectra}
\end{figure}

Figure \ref{spectra} shows the NMR spectra at several temperatures for a magnetic field perpendicular to the conduction plane. The left (right) panel shows the results for the pure ($x=0.05$) sample. Two peaks corresponding to the inner and outer sites \cite{Shimizu2006,Saito2015} appear in both spectra. For both samples, linewidth broadens with decreasing temperature and no AF ordering appears, which is consistent with the previous reports for the pure material \cite{Kawamoto2004,Shimizu2006}.

\begin{figure}[htbp]
  \includegraphics[width=8.6cm]{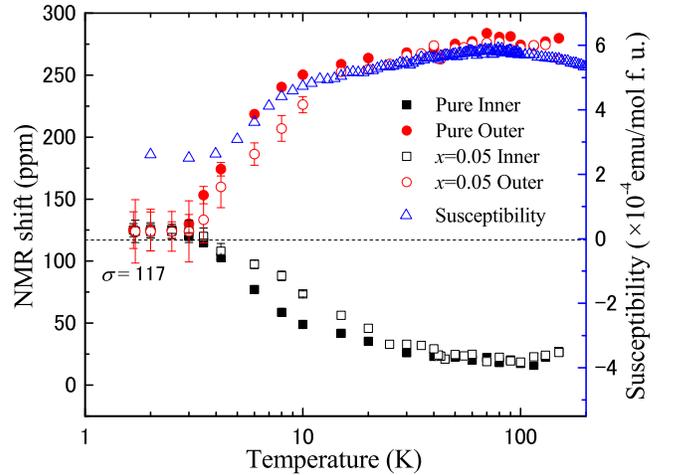}
  \caption{NMR line shift and magnetic susceptibility as a function of temperature.}\label{shift}
\end{figure}

\begin{figure}[htbp]
  \includegraphics[width=7.9cm]{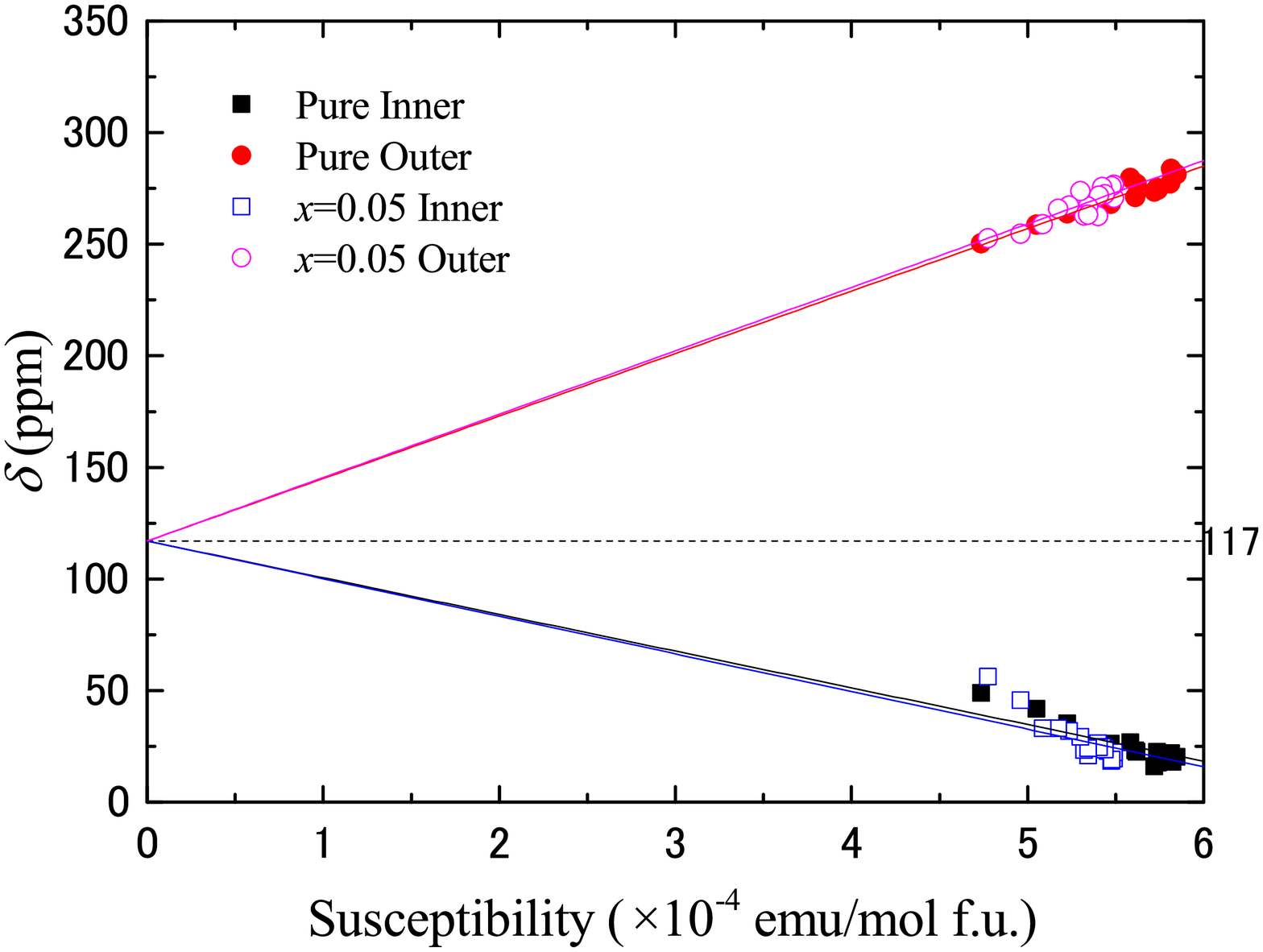}
  \caption{Temperature dependence of $\delta$-$\chi$ plot.}\label{K-x}
\end{figure}

Figure \ref{shift} shows how the NMR lines shift with temperature. Both samples showed similar line shifts that were proportional to the spin susceptibility. We obtain the hyperfine coupling constants from the $\delta$-$\chi$ plot (see Fig. \ref{K-x}). Table \ref{hyperfine} summarizes the hyperfine coupling constants, which are almost the same for both samples. The hyperfine coupling constant of the inner site is negative, whereas that of the outer site is positive; this is consistent with other $\kappa$ salts \cite{Saito2015}. As shown in Fig. \ref{susceptibility}, the finite susceptibility at 0 K seems to remain at approximately half its value at the room temperature, which is evidence of a spin liquid. Owing to slight paramagnetic impurities, the spin susceptibility depends on the subtraction of the paramagnetic impurities. NMR, however, can detect local spin susceptibility independent of paramagnetic impurities. The NMR shift $\delta$ is written as
\begin{equation}
 \delta = K + \sigma = A_{\rm hf} \chi_s + \sigma
\end{equation}
where $K$ is the Knight shift, $A_{\rm hf}$ is the hyperfine coupling constant, and $\sigma$ is the chemical shift. To discuss the spin susceptibility, the chemical shift of (BEDT-TTF)$^{+0.5}$ in this configuration was determined to be $\sigma=117$ ppm from the chemical-shift tensor of $\alpha$-(BEDT-TTF)$_2$I$_3$\cite{Kawai2009}. With decreasing temperature, the NMR shift approaches to the chemical shift.
\begin{table}
 \begin{center}
  \caption{Hyperfine coupling constants.}
   \begin{tabular}{|c|c|c|} \hline
 &value (kOe/$\mu_B$) \\ \hline
$A_{\parallel a*, \mathrm{pure\: in}}$ &-0.97  \\ \hline
$A_{\parallel a*, \mathrm{pure\: out}}$ &1.59  \\ \hline
$A_{\parallel a*, \mathrm{x=0.05\: in}}$ &-0.93  \\ \hline
$A_{\parallel a*, \mathrm{x=0.05\: out}}$ &1.56  \\ \hline
    \end{tabular}
    \label{hyperfine}
  \end{center}  
\end{table}

\begin{figure}[htbp]
  \includegraphics[width=9cm]{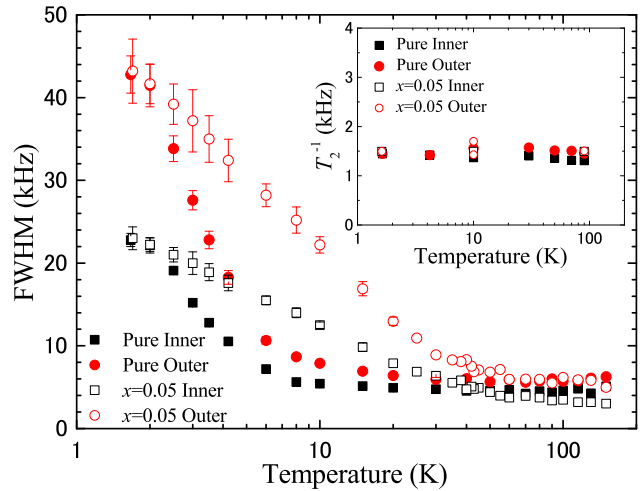}
  \caption{Temperature dependence of FWHM of NMR lines. Inset shows temperature dependence of $T_2^{-1}$.}\label{FWHM}
\end{figure}

Figure \ref{FWHM} shows the temperature dependence of the full width at half maximum (FWHM) of the NMR lines and the inset shows the temperature dependence of $T_2^{-1}$ for the inner and outer sites of both samples. For Fourier transform NMR, the linewidth $\Delta \omega$ of the spectrum is generally described as
\begin{equation}
\Delta \omega = \frac{2 \pi}{T_2}+\gamma_{\rm I} \Delta H \label{T2FWHM}
\end{equation}
where $\gamma_{\rm I}$ is the nuclear gyromagnetic ratio and $\Delta H$ is the inhomogeneity of the local magnetic field at the corresponding nuclei. The first term is inhomogeneous width caused by the dynamics and the second term is a static inhomogeneous width caused by the inhomogeneity of the external and local magnetic field. $T_2^{-1}$ can detect slow magnetic fluctuations. In the pure sample, the NMR linewidth gradually broadens with decreasing temperature, whereas $T_2^{-1}$ remains constant, indicating that the broadening is primarily to inhomogeneous broadening \cite{Shimizu2006}. The outer site which has a larger hyperfine coupling constant, has a broader NMR line than the inner site. Below 10 K, further broadening occurs. To determine the inhomogeneous linewidth $\nu$, the natural linewidth of 2.8 kHz determined by $T_2^{-1}$ is subtracted from the FWHM (see Eq. \ref{T2FWHM}). The ratio of the inhomogeneous linewidth $\nu_{\rm out}/\nu_{\rm in}$ is approximately 2 from 60 K, where the linewidth starts to broaden, down to 10 K. This ratio is comparable to the analogous ratio of the hyperfine coupling constant $|A_{\rm out}/A_{\rm in}|=1.64$. Thus, the linewidth broadening correlates with the hyperfine coupling constants, suggesting that the spin density on molecules is inhomogeneous. In other words, $\Delta K = A \Delta \chi$.

For the $x=0.05$ sample, the line broadening is detected, which is greater than that for the pure sample from 60 to 15 K. As $T_2^{-1}$ for the $x=0.05$ sample is independent of temperature, the increase in linewidths is attributed to an enhanced static inhomogeneity, as for the pure sample. The ratio of the inhomogeneous NMR linewidth $\nu_{\rm out}/\nu_{\rm in}\simeq 2$ resembles that of the hyperfine coupling constant $|A_{\rm out}/A_{\rm in}|=1.75$. This suggests the impurity substitution enhances the disorder.

Next, we focus on the anomalous increase of the FWHM of the NMR lines in the low-temperature region. The NMR linewidth of the spin singlet state is expected to be narrow. However, the FWHM does not decrease at low temperature. Recently, $\mu$SR measurements by Nakajima \textit{et al}., suggested that spins paramagnetically fluctuate in zero magnetic field, and the microscopic phase separates into the singlet phase and the paramagnetic phase below 3 K \cite{Nakajima2012}. To explain these results, Nakajima \textit{et al}., invoked microscopic paramagnetic islands surrounded by a singlet sea with a finite gap. As the volume fraction of the singlet sea is much greater than that of the paramagnetic area, the paramagnetic spin behaves as a magnetic impurity, contributing only to the linewidth of the nonmagnetic NMR spectrum. Thus, this picture is consistent with the small Knight shift with a broad linewidth. Therefore, the finite susceptibility in the static susceptibility measurements and proportionality of the linewidth to magnetic field at low temperature \cite{Shimizu2006} might be the consequence of this paramagnetic component.

For $x=0.05$ sample, the FWHM also rapidly increases below 15 K. One possible explanation of this result is that the impurity substitution induces the AF moment; however, this can be ruled out becuase if impurities induce the AF moment, the FWHM for $x=0.05$ would be broader than that for the pure sample. However, the FWHM is almost the same for both samples at 1.7 K, in contrast to what happens above 15 K. Conversely, the paramagnetic island proposed by $\mu$SR \cite{Nakajima2012} is indepedent of nonmagnetic impurities such as the BEDT-STF, so the paramagnetic islands are independent of the BEDT-STF impurities, resulting in a same-amplitude inhomogeneous dipole field in both samples, and translating into the same broadening in both samples. 

\subsection{Spin-lattice relaxation rate}

\begin{figure}[htbp]
  \includegraphics[width=7.9cm]{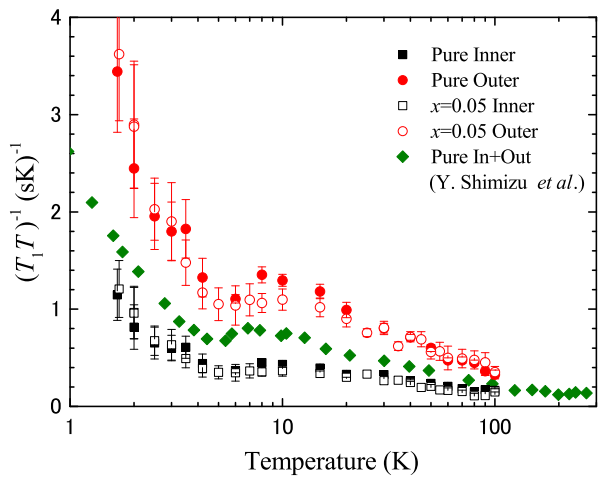}
  \caption{Temperature dependence of $(T_1T)^{-1}$ for pure sample and $x=0.05$ sample.}\label{T1T}
\end{figure}

The Spin-lattice relaxation rate $1/T_1$ probes spin fluctuation, which is written as
\begin{eqnarray}
 \frac{1}{T_{1}T} = \frac{2\gamma^2 _{I}k_{\rm B}}{ \left( \gamma _{\rm e} \hbar \right)^2 } \sum _{q} \left( A_{q}A_{-q} \right) \frac{\chi _{q} ''(\omega )}{\omega }.
\end{eqnarray}
Here, $\gamma_{\rm e}$ is the electron gyromagnetic ratio, $A$ is the hyperfine coupling constants between electron and nucleus, and ${\chi _{q} ''(\omega )}$ is the imaginary part of the dynamic susceptibility at wave vector $q$.
Figure \ref{T1T} compares the temperature dependence of $(T_1T)^{-1}$ of both samples with the result from a previous measurement of a double $^{13}$C enriched sample under the same configuration \cite{Shimizu2006}. We determined $T_1^{-1}$ by separately fitting a single exponential to the inner and outer sites, or by fitting the two-exponential models to the sum of the spectral intensity below 20 K by using the ratio, $T_{1,\mathrm{inner}}/T_{1,\mathrm{outer}}=3$, which was determined at 100 K. This value is typical of $\kappa$-(BEDT-TTF)$_2X$ salts.

As temperature decreases, the quantity $(T_1T)^{-1}$ increases and broadly peaks near 8 K. However, note that recovery curves for both samples deviate from the exponential fits below 6 K, which suggests distribution of $T_1^{-1}$. This distribution corresponds to the anomalous broadening below 8 K. Using $(T_1T)^{-1}$ of the pure sample reproduces the results of previous work \cite{Shimizu2006}. On the basis of the ratio $U/W$ of (CN)$_3$ salt, a AF transition temperature $T_{\rm N}$ comparable to that of Cl salt is expected. The suppression of the AF transition indicates the characteristics of the geometrical frustration below the temperature $T_{\rm N}=27$ K \cite{Miyagawa1995} of the Cl salt. However, the temperature dependence of $(T_1T)^{-1}$ for $x=0.05$ sample is quantitatively similar to that of the pure sample, indicating a lack of the geometrical frustration.

\subsection{NMR of impurity site}

Impurity site in frustrated spin systems are expected to behave differently than bulk site. To verify geometrical frustration, applying NMR to the impurity sites provides useful information. However, NMR spectra affected by impurities has been discussed on the basis of using long tails of spectrum line on the bulk site. Although using NMR to detect the impurity-site resonances is difficult in frustrated spin systems, we directly observed $^{13}$C-NMR on impurity sites using $^{13}$C-enriched BEDT-STF molecules.

\begin{figure}[htbp]
  \includegraphics[width=7cm]{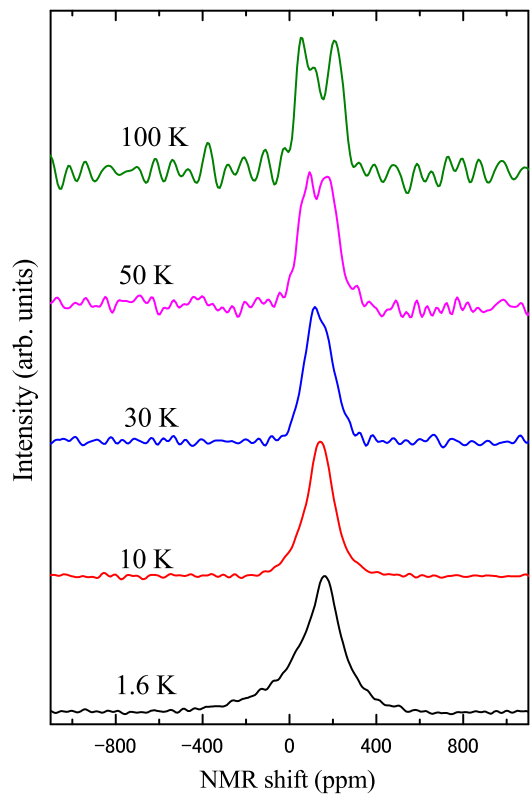}
  \caption{NMR spectra at several temperature of impurity sites.}\label{spectraimp}
\end{figure}

Figure \ref{spectraimp} shows NMR spectra from impurity sites at several temperatures. Two peaks are observed at 100 K. The NMR shift is less than that of the bulk sites, indicating a small local spin susceptibility. Linewidth broadening occurs at low temperature and no AF ordering appears.

\begin{figure}[htbp]
  \includegraphics[width=7.9cm]{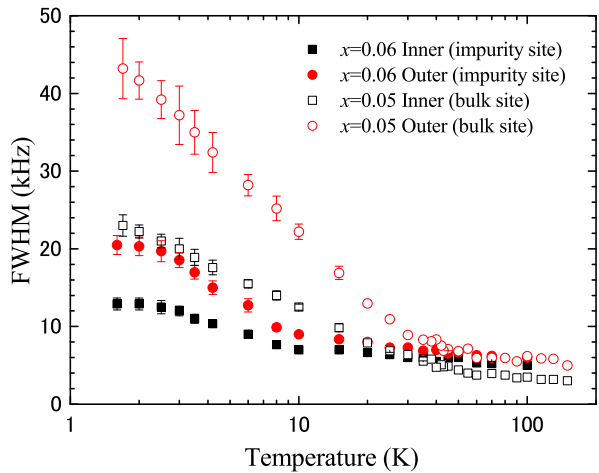}
  \caption{FWHM of NMR line for impurity and bulk sites as a function of temperature.}\label{FWHMimp}
\end{figure}

In frustrated spin systems, impurities can induce local staggered moments and line broadening is expected owing to staggered spin density oscillations, as expected from the AF character of magnetic correlations \cite{Takigawa1997,Fujiwara1998}. Figure \ref{FWHMimp} shows the temperature dependence of FWHM of the NMR lines from impurity sites as a function of temperature. The FWHM of the NMR line of the impurity site at 1.7 K is less than that of bulk sites, indicating that the impurities do not induce staggered moments.

\begin{figure}[tbp]
  \includegraphics[width=7.9cm]{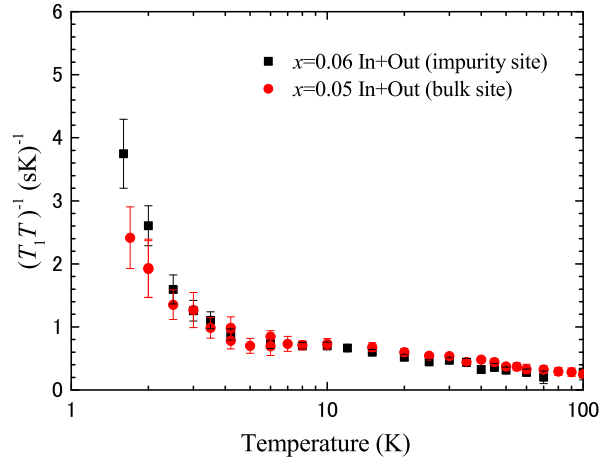}
  \caption{Temperature dependence of $(T_1T)^{-1}$ for impurity and bulk sites. $(T_1T)^{-1}$ for the impurity site is scaled by a factor of two.}\label{T1Timp}
\end{figure}

Figure \ref{T1Timp} shows $(T_1T)^{-1}$ for the bulk and $(T_1T)^{-1}$ scaled for the impurity sites, both as functions of temperature. Each curve is the mean value of the associated inner and outer sites. Scaling $(T_1T)^{-1}$ for impurity sites by a factor of two matches well with the results for $(T_1T)^{-1}$ for the bulk sites. In Fermi liquid theory, $(T_1T)^{-1}$ is proportional to the square of the product of the hyperfine coupling constant $A_{\rm hf}$ and the density of state $N(E_{\rm F})$ at the Fermi energy:
\begin{equation}
\cfrac{1}{T_1T} = \cfrac{\pi k_B}{\hbar} A_{\rm hf}^2 N^2(E_{\rm F}).
\end{equation}
Here, $N(E_{\rm F})$ corresponds to the local spin susceptibility. The temperature dependence of $(T_1T)^{-1}$ for impurity sites is similar to that for bulk sites, and the scale factor indicates a small local susceptibility compared with that of the bulk sites. The qualitatively similar behavior of both sites suggests that the electronic properties at bulk and impurity sites are described by one-fluid model. Note that recovery curves deviate from the exponential fits below 8 K, which suggests distribution of $T_1^{-1}$. The NMR spectra from impurity sites are similar to those from bulk sites, indicating that the magnetic behavior of (CN)$_3$ salt is not due to the geometrical frustration. Instead, the suppression of the AF transition and the electronic state of (CN)$_3$ salt may be due to disorder and electronic correlation, in addition to the ideal geometrical frustration, as theoretically suggested \cite{Byczuk2005,Aguiar2009,Shinaoka2009}.

\section{Conclusion}
We investigated the conductivity and the magnetic properties of (CN)$_3$ salt by artificially distorting its triangular lattice by substitution of BEDT-STF, which introduces disorder by modifying the transfer integrals among the BEDT-STF molecules.

Temperature dependence of conductivity of a sample with impurity fraction $x=0.05$ is not much changed above 200 K, indicating that the substitution enhances conductivity and leads to a crossover temperature $T_{\rm cross}$ that is greater than that of a pure sample. At low temperature, the $x=0.05$ sample exhibits VRH-like conductivity, suggesting that no intrinsic gap opens and that the sample is not a Mott insulator like the pure sample \cite{Pinteric2014}.

The NMR spectra reveal no magnetic ordering, and the spin susceptibility approaches to the chemical shift below 3 K in both samples. These results are inconsistent with finite susceptibility from the static magnetic measurement at low temperature. From 60 to 15 K, the NMR linewidth of the $x=0.05$ sample becomes broader than that of the pure sample, indicating the BEDT-STF substitution enhances disorder. The temperature dependence of $(T_1T)^{-1}$ for both samples is quantitatively similar. Moreover, the temperature dependence of $(T_1T)^{-1}$ for impurity sites is similar, suggesting that the characteristics of $(T_1T)^{-1}$ are not due to geometrical frustration. NMR spectra from impurity sites suggest a decrease in local spin susceptibility and that no staggered moments are induced. Thus, the results indicate that the static and dynamic susceptibility do not change, even at temperatures two orders of magnitude less than the exchange interaction $J\simeq$ 250 K. These results are in stark contrast to the expected effect of the substitution on conductivity. Thus, we suggest that the electronic state of (CN)$_3$ salt is not only due to the ideal geometrical frustration but also due to the disorder.

\section*{ACKNOWLEDGMENTS}
We thank M. Matsumoto for sample preparation.

\end{document}